\documentclass[twocolumn, times, twocolappendix]{aastex63}

\newcommand{\oi}{[\ion{O}{1}] }
\newcommand{\cii}{[\ion{C}{2}] }
\newcommand{\nii}{[\ion{N}{2}] }
\newcommand{\hii}{\ion{H}{2} }
\newcommand{\um}{$\mu$m\,}
\usepackage{threeparttable}

\received{}
\revised{}
\accepted{}
\submitjournal{}

\shorttitle{Dense, warm neutral gas in BR1202-0725 system}
\shortauthors{Minju M. Lee et al.}
\defcitealias{minju2019b}{Paper I}

\begin{document}
\title{Dense and warm neutral gas in BR1202-0725 at z = 4.7 as traced by the  [O I] 145 $\mu$m line}

\correspondingauthor{Minju M. Lee}
\email{minju@mpe.mpg.de}

\author[0000-0002-2419-3068]{Minju M. Lee}
\affiliation{Max-Planck-Institut f\"{u}r Extraterrestrische Physik (MPE), Giessenbachstr. 1, D-85748 Garching, Germnay}

\author[0000-0002-7402-5441]{Tohru Nagao}
\affiliation{Graduate School of Science and Engineering, Ehime University, 2-5 Bunkyo-cho, Matsuyama 790-8577, Japan}
\author[0000-0002-6637-3315]{Carlos De Breuck}
\affiliation{European Southern Observatory, Karl Schwarzschild Stra{\ss}e 2, 85748 Garching, Germany}

\author[0000-0002-6719-380X]{Stefano Carniani}
\affiliation{Scuola Normale Superiore, Piazza dei Cavalieri 7, I-56126 Pisa, Italy}

\author[0000-0002-5281-1417]{Giovanni Cresci}
\affiliation{INAF -- Osservatorio Astrofisico di Arcetri, Largo E. Fermi 5, I-20125 Firenze, Italy}

\author[0000-0001-6469-8725]{Bunyo Hatsukade}
\affiliation{Institute of Astronomy, Graduate School of Science, The University of Tokyo, 2-21-1 Osawa, Mitaka, Tokyo 181-0015, Japan}

\author[0000-0002-8049-7525]{Ryohei Kawabe}
\affiliation{National Observatory of Japan, 2-21-1 Osawa, Mitaka, Tokyo 181-8588, Japan}
\affiliation{SOKENDAI (The Graduate University for Advanced Studies), 2-21-1 Osawa, Mitaka, Tokyo 181-8588, Japan}

\author[0000-0002-4052-2394]{Kotaro Kohno}
\affiliation{Institute of Astronomy, Graduate School of Science, The University of Tokyo, 2-21-1 Osawa, Mitaka, Tokyo 181-0015, Japan}
\affiliation{Research Center for the Early Universe, The University of Tokyo, 7-3-1 Hongo, Bunkyo, Tokyo 113-0033, Japan}

\author[0000-0002-4985-3819]{Roberto Maiolino}
\affiliation{Cavendish Laboratory, University of Cambridge, 19 J. J. Thomson Avenue, Cambridge CB3 0HE, UK}
\affiliation{Kavli Institute for Cosmology, University of Cambridge, Madingley Road, Cambridge CB3 0HA, UK}

\author[0000-0002-4803-2381]{Fillipo Mannucci}
\affiliation{INAF -- Osservatorio Astrofisico di Arcetri, Largo E. Fermi 5, I-20125 Firenze, Italy}

\author[0000-0002-9889-4238]{Alessandro Marconi}
\affiliation{INAF -- Osservatorio Astrofisico di Arcetri, Largo E. Fermi 5, I-20125 Firenze, Italy}
\affiliation{Dipartimento di Fisica e Astronomia, Universit\'{a} degli Studi di Firenze, Via G. Sansone 1, I-50019 Sesto F.no, Firenze, Italy}

\author[0000-0002-6939-0372]{Kouichiro Nakanishi}
\affiliation{National Observatory of Japan, 2-21-1 Osawa, Mitaka, Tokyo 181-8588, Japan}
\affiliation{SOKENDAI (The Graduate University for Advanced Studies), 2-21-1 Osawa, Mitaka, Tokyo 181-8588, Japan}

\author[0000-0001-6162-3023]{Paulina Troncoso}
\affiliation{Escuela de Ingenier\'ia, Universidad Central de Chile, Avenida Francisco de Aguirre 0405, 171-0614, La Serena, Coquimbo, Chile}

\author[0000-0003-1937-0573]{Hideki Umehata}
\affiliation{RIKEN Cluster for Pioneering Research, 2-1 Hirosawa, Wako, Saitama 351-0198, Japan}
\affiliation{Institute of Astronomy, Graduate School of Science, The University of Tokyo, 2-21-1 Osawa, Mitaka, Tokyo 181-0015, Japan}

%\collaboration{1}{(AAS Journals Data Scientists collaboration)}
%\nocollaboration{1}
%\collaboration{1}{(LaTeX collaboration)}



\begin{abstract}
We report the detection of \oi 145.5$\mu$m in the BR 1202-0725 system, a compact group at $z=4.7$ consisting of a quasar (QSO), a submillimeter-bright galaxy (SMG), and three faint Ly$\alpha$ emitters.
By taking into account the previous detections and upper limits, the \oi/\cii line ratios of the now five known high-$z$ galaxies are higher than or on the high-end of the observed values in local galaxies (\oi/\cii$\gtrsim 0.13$). 
The high \oi/\cii ratios and the joint analysis with previous detection of \nii lines for both of the QSO and the SMG suggest the presence of warm and dense neutral gas in these highly star-forming galaxies.
This is further supported by new CO~(12--11) line detections and a comparison with cosmological simulations.
There is a possible positive correlation between the \nii122/205 line ratio and the \oi/\cii ratio when all local and high-z sources are taken into account, indicating that the denser the ionized gas, the denser and warmer the neutral gas (or vice versa).
The detection of the \oi line in the BR1202-0725 system with a relatively short amount of integration with ALMA demonstrates the great potential of this line as a dense gas tracer for high-$z$ galaxies. 
\end{abstract}



\section{Introduction} \label{sec:intro}

One of the key questions in modern astrophysics is to understand the physical processes that govern the star formation and galaxy assembly in the early universe.
Compared to typical star-forming galaxies on the main-sequence (e.g., \citealt{Noeske2007, Elbaz2007, Speagle2014}), QSOs and dusty star-forming galaxies -- we will call the latter population as submillimeter-bright galaxies, hereafter SMGs\footnote{A recently used term is dusty star-forming galaxies (DSFGs) but we decided to use the conventional term SMG, here, because the name has been used for BR1202-0725 for a long time.}, in the generally accepted view -- release enormous amount of energies coming from active black-hole accretion and/or star formation.
In this paper, we refer to main-sequence (MS) galaxies as those within $\pm0.2$ dex from the definition of \citet{Speagle2014}, and starburst galaxies as galaxies at least 3$\sigma$ above the main-sequence (i.e., $\log{\Delta {\rm MS}}> 0.6$). 
While QSOs and SMGs were discovered by different methods, both populations often represent star-bursting galaxies.
Of particular interest is understanding how their star formation activities are regulated, their comparison with normal populations (i.e., main-sequence galaxies), and how they impact surroundings.

Oxygen is the third most abundant element in the universe. 
The neutral oxygen has a ionization potential of 13.62 eV, which is close to that of hydrogen (13.59 eV), so the \oi emission line arises mostly from neutral regions. 
Fine structure lines of oxygen serve as one of the main coolants of the interstellar medium (ISM) at far-infrared (FIR) (e.g, \citealt{Rosenberg2015}).
With a critical density of $\sim10^5$ cm$^{-3}$, \oi line traces much denser ISM than the \cii emission line.
By constraining the physical properties (mainly the strength of the radiation field and density), one can infer the dense gas distributions where star-forming activity would take in place (e.g., \citealt{Malhotra2001}).
This will essentially lead us to understand galaxy formation and evolution.

The first detection of \oi ($^3P_0-^3P_1$) at 145.5 \um (hereafter, \oi $_{145}$) was reported in 1983 by \citet{Stacey1983}, but the number of galaxies detected in this lower level transition was limited largely due to its fainter nature relative to the higher transition of \oi ($^3P_1-^3P_2$) at 63.2~\um (hereafter, \oi$_{63}$).
The situation was greatly improved with the advent of the Herschel space telescope which allowed studies of detailed ISM conditions of galaxies ranging from ultra-luminous infra-red galaxies (ULIRGs) to low metallicity dwarf galaxies together with other fine-structure lines (e.g., \citealt{Farrah2013, Spinoglio2015, Cormier2015,FernandezOntiveros2016, Herrera-Camus2018a}).
For high redshift galaxies ($z>3.1$), the line falls into transmission windows that are observable from ground-based telescopes.
However, less than a handful of galaxies have been observed and detected in \oi$_{145}$ with help of galaxy lensing (\citealt{Yang2019, DeBreuck2019}) or for luminous QSO-host galaxies (\citealt{Novak2019}: non-detection ; \citealt{Li2020}).

The BR1202-0725 system is a compact group at redshift $z=4.7$ consisting of a QSO, a SMG and (at least) three Lyman alpha emitters (LAEs; LAE~1, LAE~2, and LAE~3) (\citealt{Hu1996, Williams2014, Drake2020})\footnote{Recent MUSE/VLT observations suggested that Ly$\alpha$ emission from the companion dubbed LAE 2 may be part of the QSO's extended halo, though the  presence of a QSO companion, close to the LAE~2, is confirmed by the detection of dust continuum, \cii and \nii$_{205}$ (the \nii line is marginal detection with S/N$\sim$3) line emissions (\citealt{Wagg2012, Carilli2013b, Decarli2014}).}.
The BR1202-QSO and BR1202-SMG are highly star-bursting galaxies with $L_{\rm FIR}$ of $\sim10^{13} \, L_{\odot}$ (corresponding to SFR of $\sim$1000 $M_{\odot}$ yr$^{-1}$; e.g., \citealt{Omont1996, Yun2000, Iono2006}).
In our previous paper (\citealt{minju2019b}), we reported the first detection of \nii 122 \um and discussed the ionized gas density for the first time at this redshift, based on the \nii122$\mu$m/\nii205$\mu$m line ratio (hereafter, \nii$_{122/205}$).
In this following paper, we report \oi$_{145}$ line detections from both the QSO and the SMG.
By adding these two detections, the total number of the \oi$_{145}$ detection at $z>4$ has now reached five.

This work is organized as follows. In Section~\ref{sec:observations}, we explain the observations including ancillary data sets and data analyses.
In Section~\ref{sec:results}, we describe the line detection and discuss the \oi/\cii line ratio in comparison with other galaxies.
In Section~\ref{sec:discussion}, we discuss the ISM conditions inferred from the line ratio gathering other available information and summarize our findings.
We adopt a standard $\Lambda$CDM cosmology with $H_0 =70$ km s$^{-1}$ Mpc$^{-1}$ and $\Omega_m=0.3$ and Chabrier initial mass function (IMF; \citealt{Chabrier2003}).

\section{Observations and data analysis} \label{sec:observations}
\subsection{Band 7 : \oi145.5~$\mu$m observations}
The \oi 145 line observations were part of our ALMA Cycle 2 program (ID : \#2013.1.00745.S, PI : T. Nagao).
The ALMA observations used 37 antennas with the baseline length ($L_{\rm baseline}$ between 15--558 m (C34-2/1, C34-3/(4))
on 2014 December 13 and 2015 May 14 (total on-source time of T$_{\rm integ}$ = 39 minutes).

The set-up for the correlator was four spectral windows (SPW), two of which were set to each sideband, each of the SPWs with a 1.875 GHz bandwidth. 
The spectral resolution for the upper sideband was set to 
3.906 MHz ($\sim$ 3.2 km s$^{-1}$) to detect the \oi line and 7.812 MHz for the lower sideband ($\sim$ 6.7 km s$^{-1}$).
A strong quasar J1058+0133 and J1256-0547 were chosen for bandpass calibration. 
J1256-0547 was also the phase calibrator for the Band 7 observations.
Ganymede and Titan were chosen for the flux calibrator in Band 7.

We used Common Astronomy Software Applications ($\mathtt{CASA}$) (\citealt{McMullin2007}) version 4.2.2 for calibration using the pipeline script provided by the ALMA Regional Center staffs.
We then used CASA 5.6.1 version for imaging and analyzing.
Images were produced by $\mathtt{CASA}$ task, $\mathtt{tclean}$, and deconvolved down to 1$\sigma$ noise level.
The synthesized beam size with natural weighting is $0''.69\times0''.43$.
Tapered images were also created with uvtaper paramters of $0''.6, 0''.8, 1''.0$ (the corresponding synthesized beams are $0''.92\times0''.71$, $1''.05\times0''.86$, and $1''.18\times1''.03$, respectively) to check the existence of any extended emissions, especially for QSO, that could resemble the Ly$\alpha$ halo (\citealt{Drake2020}).
We first CLEANed images from $uv$ visibilities without continuum subtraction where the CLEAN masks were made based on the position of each source (i.e., QSO, SMG, LAE~1 and LAE~2) with a spectral binning of 100 km s$^{-1}$.
Continuum subtraction was applied using $\mathtt{imcontsub}$ after the full field-of-view (FoV) image was created. It was intended to obtain improved results of continuum subtraction for galaxies off from the phase center.
The continuum was subtracted with a linear fit by choosing line-free channels.
The 1$\sigma$ noise level after continuum subtraction at 100 km s$^{-1}$ is 0.40 mJy beam$^{-1}$ for natural weighting. 
For tapered images, the noise levels are 0.46, 0.51 and 0.57 mJy beam$^{-1}$ for uvtaper parameters of $0''.6$, $0''.8$ and $1''.0$, respectively, at 100 km s$^{-1}$.
Primary beam correction was also applied to get final images.

Line intensity maps were created by choosing a channel range which gives the highest peak signal-to-noise ratio (S/N). 
We measured line flux using the integrated map for each source and investigated the curve of growth as a function of aperture sizes using $\mathtt{imfit}$.
Typically, the peak S/Ns were the highest when the aperture size was set to a double the beam size.
We measured the underlying Gaussian area in the 1D spectrum as well, which gives consistent values within the errors compared to the fitted values in the 2D images. 
The line widths are measured using the same aperture size ($1''.4$) and fitted the spectrum with a single Gaussian component.
We further investigated the reliability image-based continuum subtraction by performing continuum subtraction in a 1D spectrum (for each source) adopting the same aperture size, which also gave consistent values of fluxes (i.e., 1D Gaussian area) and line widths within the fitting errors. 
As for the final measurements for fluxes and line widths, we used the $0''.6$-tapered image cube and took the aperture of $1''.4$.

\subsection{Ancillary data sets of CO~(12--11), HCN~(6--5), and HCO$^{+}$(6--5) line observations}
To demonstrate a supportive argument for the \oi detections, we take two more data sets that our team were awarded using ALMA as PI programs.
One is from the same ALMA Cycle 2 program (ID: \#2013.1.00745.S, PI : T. Nagao) where the CO~(12--11) (Band~6) line was also targeted.
This ALMA program was designed to detect multiple fine-structure lines from the BR1202-0725 system, including the \oi$_{145}$ line, as described above, and two \nii fine-structure lines at 122~$\mu$m and 205~$\mu$m.
The Band~6 observations targeted the \nii~205$\mu$m and CO~(12--11) line at the same time at the upper and lower side band each. See the details of the observational summary presented in \citet{minju2019b}. 
The CO~(12--11) line is detected (see Section~\ref{sec:others}) from the SMG and the QSO and we present the line profiles and maps in Appendix~\ref{app:co1211}.

The other one is the Band~3 observations (ID: \#2013.1.00259.S, PI : M. Lee) which was only partially executed (20\%, $\approx 97$ mins) out of 8.4 hrs requested.
In this program, we aimed at detecting two dense gas tracers of HCN~(6--5) and HCO$^{+}$(6--5) to constrain the dense gas fraction and the lines are not detected for this partial execution (see Appendix~\ref{app:hcn}).
The critical densities of the high-$J$ HCN and HCO$^{+}$ transitions are two orders of magnitude higher that of \oi.
Still, the non-detection of the lines alternatively demonstrate the strength of the \oi line as a dense gas ($\gtrsim 10^5$ cm$^{-3}$) tracer that was detected with less than a half of the time executed for the Band~3 observations.

\section{Results} \label{sec:results}
\subsection{Detection of \oi$_{145}$ and line properties}

\begin{figure}[tb]
\centering
\includegraphics[width=0.48\textwidth, bb=0 0 1000 1000]{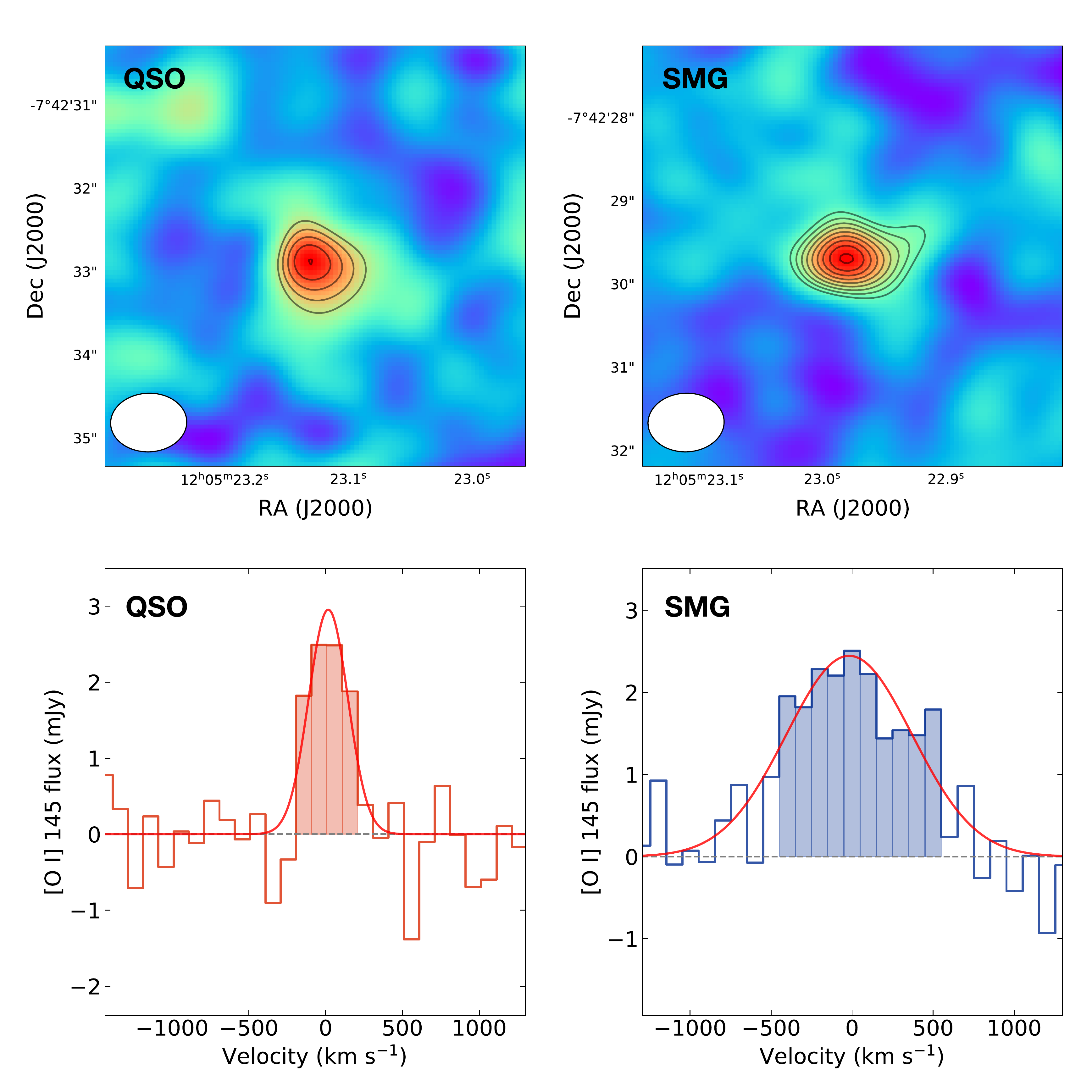}%, bb = 0 0 1024 768 ,
\caption{Upper panel: \oi$_{145}$ line maps for QSO (left) and SMG (right). The contours are drawn starting from 4$\sigma$ in steps of 1$\sigma$ up to 12$\sigma$, i.e., [4,5,6, ..., 12]$\sigma$. Negative component contours at -4$\sigma$ is also drawn as dashed lines which do not exist nearby the sources. On the bottom left on each panel, we show the synthesized beam of the final image which is $0''.92 \times 0''.71$ (tapered image with uvtaper=$0''.6$). The panel size is $5''$ in width. Bottom panel: the \oi$_{145}$ line spectrum using the same image in the upper panel for the QSO (left) and SMG (right) by taking an aperture of $1''.4$. Shaded area is the velocity range to create the line intensity map. The spectral resolution is set to 100 km s$^{-1}$. The red curve in each panel shows the best-fit Gaussian fit.\label{fig:spectrum}}
\end{figure}

\begin{table*}[bt]
\caption{Observational parameters of \oi$_{145}$ and CO~(12--11) in BR1202-0725}
\begin{center}
\begin{tabular}{cccccc}
\hline \hline
	\oi$_{145}$		& QSO 			& 	SMG  & LAE1 & LAE2\tnote[$^{a}$ & LAE3\\ \hline
$F_{\rm line}$ [Jy km s$^{-1}$] 		&	$1.19\pm 0.17$ &	$2.49\pm 0.24$		& $<0.12$\tnote[$^{b}$ & $<0.79$\tnote[$^{b}$ & $<0.50$\tnote[$^{b}$\\
FWHM [km s$^{-1}$]		& $301 \pm 139$	&	$916 \pm 225$		& 56\tnote[$^{c}$	& 338\tnote[$^{c}$ & ...\\
$L_{\rm line}$ [$10^{9} L_{\odot}$]			&	$0.84 \pm 0.12$	& 	$1.75 \pm 0.17$ &...&...&...	\\
\hline\hline
	CO~(12--11)		& QSO 			& 	SMG  & LAE1 & LAE2\tnote[$^{a}$ & LAE3 \\ \hline
$F_{\rm line}$ [Jy km s$^{-1}$] 		&	$1.03\pm 0.05$ &  $2.03\pm 0.09$ & $<0.05$\tnote[$^{d}$ &$<0.32$\tnote[$^{d}$  &$<0.19$\tnote[$^{d}$ \\
FWHM [km s$^{-1}$]		& $373 \pm 88$	&	$1058 \pm 224$ & 56\tnote[$^{c}$	& 338\tnote[$^{c}$ & ...\\
$L_{\rm line}$ [$10^{9} L_{\odot}$]			&	$\replaced{0.51\pm0.02}{0.49\pm0.02}$	& 	$\replaced{1.01\pm0.02}{0.96\pm0.04}$	& ...&...&...\\
\hline
\end{tabular}
\begin{tablenotes}
\item{$^{a}$\citet{Drake2020} concluded that ``LAE2" is not the powering source of the Ly$\alpha$ emission seen the HST~775W map. Instead, the HST emission is a stellar component passing through the QSO halo and is outshone by the halo. However, a QSO companion close to LAE~2 is confirmed by dust, \cii, and (marginally) \nii emissions (\citealt{Wagg2012, Carilli2013b, Decarli2014}), so we use the name ``LAE2" to indicate this companion.}
\item{$^{b}$ 3$\sigma$ upper limit for an aperture of 1''.4 in the tapered image with uvtaper=$0''.6$. The 3$\sigma$ limit is corresponding to a Gaussian area assuming the FWHM of the lines and using the average channel noise. We adopt the FWHM values to be the same as the \cii from \citet{Carilli2013b} for LAE~1 and LAE~2. For LAE~3, we assumed FWHM = 200 km s$^{-1}$, considering the reported FWHM in other literature for low-mass galaxies (e.g., \citealt{Pavesi2018, Bethermin2020}). All assumed FWHM values for LAEs are narrower than those of the Ly$\alpha$ emissions reported in \citet{Drake2020}. The noise is calculated based on the tapered cube (uvtaper = $0''.6$) with a spectral resolution of 100 km s$^{-1}$.}
\item{$^{c}$ From \cii observations in \citet{Carilli2013b}.}
\item{$^{d}$ 3$\sigma$ upper limit for an aperture of 2''.0. We assumed the same FWHM values that were assumed in the \oi line esimates. The noise is calculated based on a cube with a spectral resolution of 50 km s$^{-1}$.}
  \end{tablenotes}
\end{center}
\label{tab:lineprop}
\end{table*}%

\begin{figure*}[tb]
\centering
\includegraphics[width=0.80\textwidth, bb=0 0 950 650]{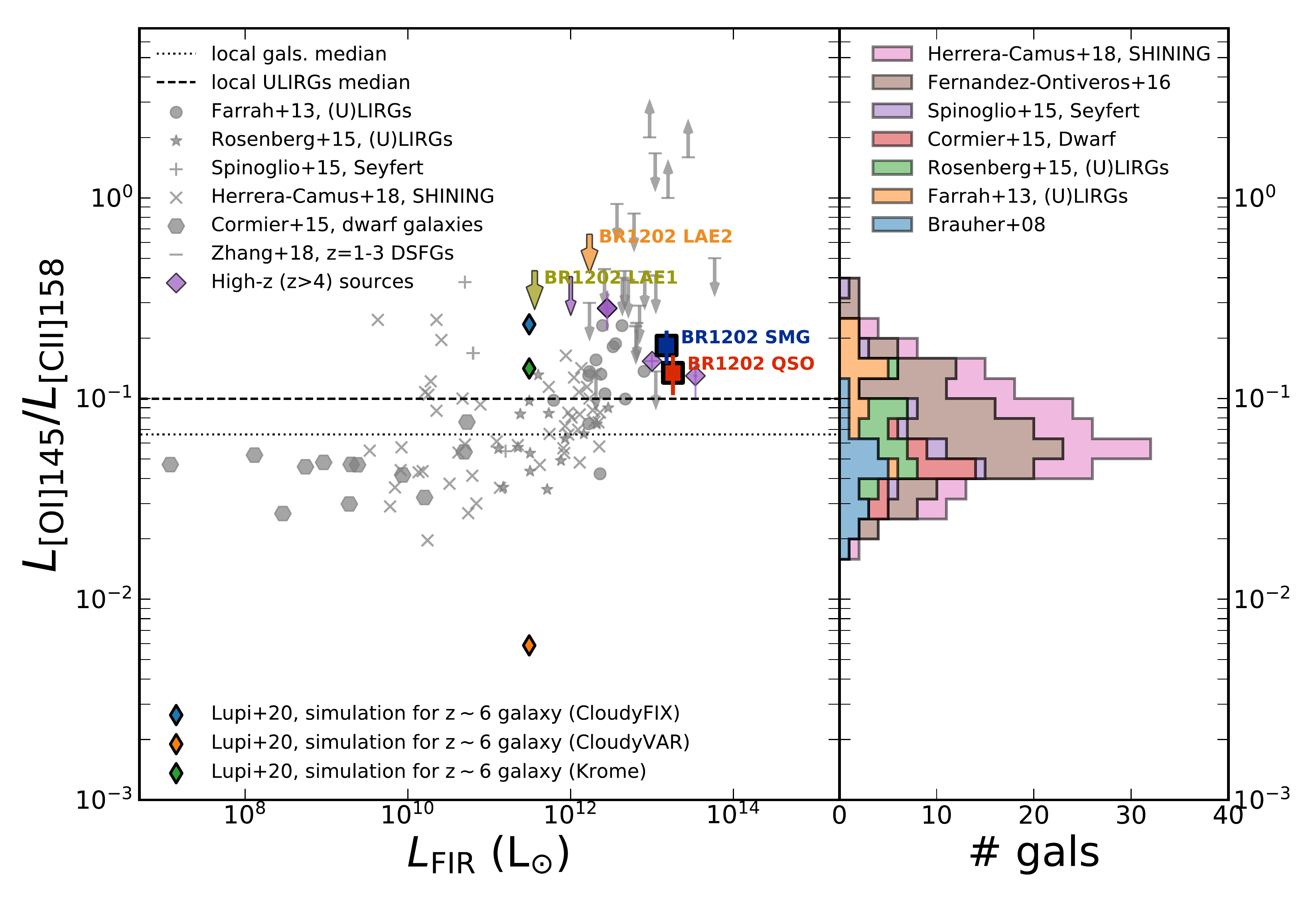}
\caption{Left: Luminosity line ratios of \oi$_{145}$/\cii$_{158}$ as a function of FIR luminosity. Four galaxies associated in the BR1202-0725 system (SMG, QSO, LAE~1, and LAE~2) are plotted with the labels for the detection or the 3$\sigma$ upper limit constraints. For local studies, we plot \citet{Farrah2013}, \citet{Cormier2015}, \citet{Rosenberg2015}, \citet{Spinoglio2015} and \citet{Herrera-Camus2018a} for which $L_{\rm FIR}$ values are available, ranging from local dwarfs to ULIRGs and Seyfert galaxies. 
We used the latest measurement when the same galaxies are listed in different literature.
As no detection data points are available for $z=1-3$ SMGs (and normal galaxies), we plot upper and lower limits for SMGs from \citet{Zhang2018}. 
Higher-$z$ ($z>4$) sources (SMGs and QSOs with high L$_{\rm FIR}$) are plotted as purple diamonds (\citealt{DeBreuck2019, Yang2019, Li2020}) and as an arrow for an upper limit from \citet{Novak2019}. In addition to observational data points, three additional data points (thin diamonds) are also plotted from the hydrodynamical simulations reported in \citet{Lupi2020} using different model assumptions (see text for details).
We show the median values of local galaxies and ULIRGs as dotted and dashed lines, respectively.
Right: Stacked histogram of luminosity line ratios of \oi$_{145}$/\cii$_{158}$ for local galaxies. 
The bins for the histogram are set to have linear steps in log space. 
For the histogram, we include local studies from \citet{Brauher2008}, \citet{Farrah2013} ((U)LIRGs),  \citet{Rosenberg2015} ((U)LIRGs), \citet{Cormier2015} (dwarf), \citet{Spinoglio2015} (Seyfert), \citet{FernandezOntiveros2016}, and \citet{Herrera-Camus2018a}. 
The latest measurements are plotted when the same galaxies are listed in different literature.\label{fig:otoc}}
\end{figure*}

The bottom panels of Figure~\ref{fig:spectrum} show the line flux versus the velocity of the QSO (left) and SMG (right). 
We detect the \oi line in the QSO and SMG with a signal to noise ratio of 7 and 10, respectively.
The \oi$_{145}$ line from the SMG is detected at a higher significance level than the QSO.
The \oi$_{145}$ is not detected in LAE~1 and LAE~2 at the redshifts and the positions reported in \citet{Carilli2013b}. 
The positions in \citet{Carilli2013b} are from the submillimeter continuum (SMG, QSO, and LAE~2) and \cii (LAE~1) emissions. 
The line search for every target was based on checking the peak S/N in a fixed aperture by varying the integrating range within the full velocity coverage of the upper sideband. 
The searching area is fixed to the beam size for all but LAE~1. 
As the optical position (of the redshifted Ly$\alpha$ emission) for LAE~1 is offset from the \cii position by $\approx0''.6$ (\citealt{Drake2020}), we search for a line detection around a $1''.0$-radius circular region from the [CII] peak to check if there is any emission associated to the Ly$\alpha$ emission instead of the \cii. 
There is no significant detection signature for the LAE~1 above 3$\sigma$.
We summarize the line properties in Table~\ref{tab:lineprop} for the \oi line.

The line widths (full-width half-maximum, FWHM) are $916\pm225$ and $301\pm139$ km s$^{-1}$ for the SMG and the QSO, respectively.
They are consistent with those of \cii ($\sim$700 km s$^{-1}$ (SMG), $\sim$300 km s$^{-1}$ (QSO); \citealt{Wagg2012, Carilli2013b, Carniani2013}) within uncertainties, but there is a hint of a broader \oi$_{145}$ line width than the \cii$_{158}$ for the SMG.
The \nii fine-structure lines in our previous work (\citealt{minju2019b}) also showed such a signature (i.e., a broader line width) for the SMG with FWHM $\sim 900$ km s$^{-1}$.
As for the QSO, the \oi$_{145}$ line widths are consistent with those of CO, \cii$_{158}$ and \nii$_{205}$, but \nii$_{122}$ reported in the literature (\citealt{Salome2012, Wagg2012, minju2019b}).
Recent study of star-forming galaxies at $z\sim6$ have shown that different emission lines trace different components of galaxies (e.g., \citealt{Carniani2017, Carniani2020}). However, the spatial resolution of current observations is not sufficient to spatially resolve the emission of the FIR lines in our galaxies. 
Future high-resolution observations will be able to reveal the origin of the discrepancy in line width and whether the FIR lines are tracing distinct regions of the galaxies.
Having the resolution limit, we regard the lines as originating from the same regions, at least globally.
The derived \oi$_{145}$ line luminosities are $(1.75\pm0.17)\times 10^9\, L_{\odot}$ and $(0.84\pm0.12)\times 10^9\, L_{\odot}$ for the SMG and the QSO, respectively.

\subsection{[O I]/[C II] Line ratio}
The derived line luminosity ratios are $L_{\rm [O I]145}$/$L_{\rm [C II] 158}$ = $0.18  \pm 0.03$ and $0.13 \pm 0.03$ for the SMG and the QSO, respectively.
The upper limits on the line ratio for LAE~1 and LAE~2 are $<0.43$ and $<0.66$, respectively, using the linewidth constraints in the literature (see Table~\ref{tab:lineprop} footnote).
In the right panel of Figure~\ref{fig:otoc}, we plot a stacked histogram of the \oi$_{145}$/\cii$_{158}$ line ratios of local galaxies. 
It includes various types of galaxies from dwarf galaxies to ULIRGs and Seyfert. 
They are obtained from \citet{Brauher2008}, \citet{Farrah2013},  \citet{Rosenberg2015}, \citet{Cormier2015}, \citet{Spinoglio2015}, \citet{FernandezOntiveros2016}, and \citet{Herrera-Camus2018a}.
When there are different flux measurements on the same galaxies from the literature we plot the most recent ones.
Out of 187 local galaxies considered, the median (mean) value of $L_{\rm [O I]145}$/$L_{\rm [C II] 158}$ is 0.07 (0.08) with a standard deviation of 0.06: $\approx84\%$ of galaxies have the \oi-to-\cii luminosity ratio below 0.13, which is the lowest value observed in the $z>4$ galaxies. Among the local galaxies, ULIRGs (30 galaxies from \citealt{Farrah2013, Rosenberg2015, Herrera-Camus2018a}) exhibit a slightly higher median (mean) value of 0.10 (0.11) with a standard deviation 0.05, which is comparable (at 1$\sigma$) to high-z galaxies. We plot these median values on the left panel of Figure~\ref{fig:otoc}. 
Considering these, we conclude that high-$z$ SMGs and QSOs have higher \oi-to-\cii ratios compared to typical local galaxies at $\sim1\sigma$ that are consistent with the high-end values observed in local ULIRGs.

In the left panel of Figure~\ref{fig:otoc}, we show the \oi/\cii line ratio as a function of FIR luminosity for galaxies with FIR values reported. 
We used the 340 GHz flux (\citealt{Iono2006, Wagg2012, Carilli2013b}) to convert it into $L_{\rm FIR}$. For LAE~1 and LAE~2, we adopted the value reported in \citet{Carilli2013b} and the FIR luminosities are $L_{\rm FIR}<3.6\times10^{11}~L_{\odot}$ and $1.7\times10^{12}~L_{\odot}$ for LAE~1 and LAE~2, respectively.

There is a hint of the \oi/\cii ratio increase as a function of FIR luminosity ($L_{\rm FIR}$) for $\log{L_{\rm FIR}/L_{\odot}}\gtrsim10$ in log scale.
High-$z$ sources detected in both lines align well with this trend that for galaxies with higher $L_{\rm FIR}$ tend to exhibit higher \oi/\cii values within the observed scatter.
If we only consider the local galaxy studies, which will be {\it complete} and not limited by sensitivity in most cases, the positive correlation is observed for $\log{L_{\rm FIR}}\gtrsim10$ with several exceptional data points from the SHINING survey (\citealt{Herrera-Camus2018a}).
For a lower $L_{\rm FIR}$, e.g., in dwarf galaxies, the \oi/\cii ratio seems to saturate into a roughly constant value of $\sim$ 0.02-0.05, which is also observed in the histogram on the right panel of Figure~\ref{fig:otoc}.
The non-detection of \oi$_{145}$ and uncertain $L_{\rm FIR}$ in the LAEs do not put stricter constraints on the correlation between the \oi-to-\cii line ratio and $L_{\rm FIR}$.

The higher \oi-to-\cii line ratios observed in the high-$z$ galaxies may be due to their L$_{\rm FIR}$ compared to local samples. 
Meanwhile, the fact that high-$z$ galaxies exhibit a higher value than the average of ULIRG might suggest that ISM properties at fixed $L_{\rm FIR}$ evolves as a function of redshift. 
Future observations with larger number of galaxies will verify this and it will allow us to understand whether high-$z$ starburst galaxies are different from local populations.

\section{Discussion}\label{sec:discussion}
\oi$_{145}$ line originates solely from neutral regions with a critical density of $\sim10^5$ cm$^{-3}$, while the \cii line can come both from ionized and neutral regions. 
Local galaxy studies have shown that the \oi$_{63}$/\cii$_{158}$ ratio is a good tracer of the photo-dissociation region (PDR, or neutral gas) density, once the \cii emission from ionized regions is subtracted, while the \oi$_{63}$/\oi$_{145}$ is a tracer of the neutral gas temperature for a range between 100 and 400 K (e.g., \citealt{Malhotra2001, FernandezOntiveros2016}) with the caveats of optical depth effect and self-absorption in \oi$_{63}$.
The line ratio between \oi$_{145}$ and \cii (\oi/\cii) can be therefore, used as a tracer of the gas pressure in the neutral region.
In the following, we first constrain the neutral gas fraction of the \cii emissions and then discuss the physical meanings of the observed \oi-to-\cii line ratio based on the dust temperature constraint, the detection of CO~(12--11), and comparison with a cosmological model.

\subsection{Neutral fraction of the [CII] line}
In order to infer a neutral gas fraction of the [C II] emission first, we calculate the fraction following local galaxy studies (e.g., \citealt{Croxall2017, DiazSantos2017, Herrera-Camus2018a}) using the \nii$_{205}$/\cii$_{158}$ line ratio:
\begin{equation}
f^{\rm neutral}_{\rm [C II]} =  1-  R_{\rm ion} \left( \frac{\rm [N II]_{205}}{\rm [C II]_{158}} \right)
\end{equation}
where $R_{\rm ion}$ is the \cii$_{158}$/\nii$_{205}$ luminosity ratio if the \cii$_{158}$ line is originated only from ionized regions.
\citet{Croxall2017} showed that $R_{\rm ion}$ is almost constant ranging between 2.5 and 3 for an electron density range of $n_e = [10-200]$ cm$^{-3}$ using the collision rates of \citet{Tayal2008, Tayal2011} and Galactic gas phase abundances of nitrogen (\citealt{Meyer1997}) and carbon (\citealt{Sofia2004}) (see also \citealt{Malhotra2001, Oberst2006}).
The $n_e$ values constrained in \citet{minju2019b} are $n_{\rm ion}$ = $26^{+12}_{-11}$ and $134^{+50}_{-39}$ cm$^{-3}$ for the SMG and QSO, respectively.
Therefore, it is reasonable to assume a value between 2.5 and 3 as $R_{\rm ion}$.
We use the \cii and \nii flux values reported in \citet{Wagg2012} and \citet{minju2019b}, respectively.
We note that \citet{Carniani2013} reported lower values for the \cii flux due to a smaller beam size from their differently $uv$-weighted images.
\citet{Decarli2014} reported the \nii$_{205}$ observations of this system from IRAM Plateau de Bure Interferometer (PdBI) observations. But the published flux values (or upper limits) are higher than what we obtained from our ALMA observations. 
This discrepancy can be attributed to their low S/Ns and different ways of flux measurements. As discussed in \citet{minju2019b}, our \nii205 flux is consistent with the measurement by \citet{Pavesi2016}, which is based on our data set, and further checked with other independent (ALMA) data reported in \citet{Lu2017b}.
The systematic errors from the flux measurement method could change the flux value by a factor of 3, however. As it is difficult to pin down the origin of the difference in different observations, we stick to our best measurement listed in \citet{minju2019b}, where the flux measurements are done similarly to the work presented here.

The inferred neutral fraction, $f^{\rm neutral}_{\rm [C II]}$, is 79-83\% (SMG) and 83-86\% (QSO) for the observed \nii$_{205}$/\cii$_{158}$ luminosity ratios of $14.5 \pm 1.2$(SMG) and $17.8 \pm 1.3$ (QSO).
For comparison, local studies have found that the contribution of the \cii emission originated from \hii regions to the total \cii is not dominant (less than 50\%) across a wide range of SFR density and metallicity (\citealt{Croxall2017, DiazSantos2017, Herrera-Camus2018a}).
Similarly, for the QSO and the SMG at $z=4.7$, the \cii emission is mostly coming from neutral regions rather than the ionized.
With the $f^{\rm neutral}_{\rm [C II]}$ constraints, we use the \oi$_{145}$/\cii$_{158}$ ratio as a tracer for the neutral gas density and gas temperature (thus the gas pressure), without imposing any assumptions of interstellar medium (ISM) structure.

\subsection{Dust temperature and CO~(12--11) detection}\label{sec:others}

We infer $T_{\rm dust}$ from an empirical fitting that connects the \oi/\cii$_{\rm neutral}$ line ratio to $T_{\rm dust}$ through the FIR color ($S_{63}/S_{158}$) (\citealt{DiazSantos2017}, using Equation 6 and 2 in the paper).
It gives 43 K and 54 K for the SMG and the QSO, respectively and they are well-matched with the fitted SED dust temperature in \citet{Salome2012}.
For this calculation we have assumed \oi$_{63}$/\oi$_{145}$ = 10, and the neutral fraction of the \cii$_{158}$ emission based on the \nii$_{205}$/\cii$_{158}$ line ratio as explained above.
We also note that the high neutral gas fraction of the \cii emission is also coupled with the warm dust temperature.
The \oi$_{63}$/\oi$_{145}$ ratio below 10 would imply an optically thick emission (\citealt{Tielens1985a}) and many local galaxies (except for a few extraordinary galaxies, e.g., Arp220, IRAS17208-0014 with self-absorption of \oi$_{63}$) have \oi$_{63}$/\oi$_{145}$ ratios higher than that.
For the optically thin case, the inferred dust temperature would be higher.
The detection of \oi$_{63}$ will be very challenging from ground-based telescopes for the BR1202-0725 system; the line would fall into an ALMA Band 10 coverage, where the transmission is below 0.2 at a best precipitable water vapor condition and the line \oi$_{63}$ may also suffer from self-absorption for these dusty population.
Nevertheless, the remarkable agreement between the dust temperature from different approaches strengthens the view that for both the SMG and the QSO, the observed high \oi$_{145}$/\cii$_{158}$ ratios are closely connected to their warm dust temperatures and hence high kinetic gas temperatures.

\begin{figure}[tb]
\centering
\includegraphics[width=0.48\textwidth, bb=0 0 1000 1000]{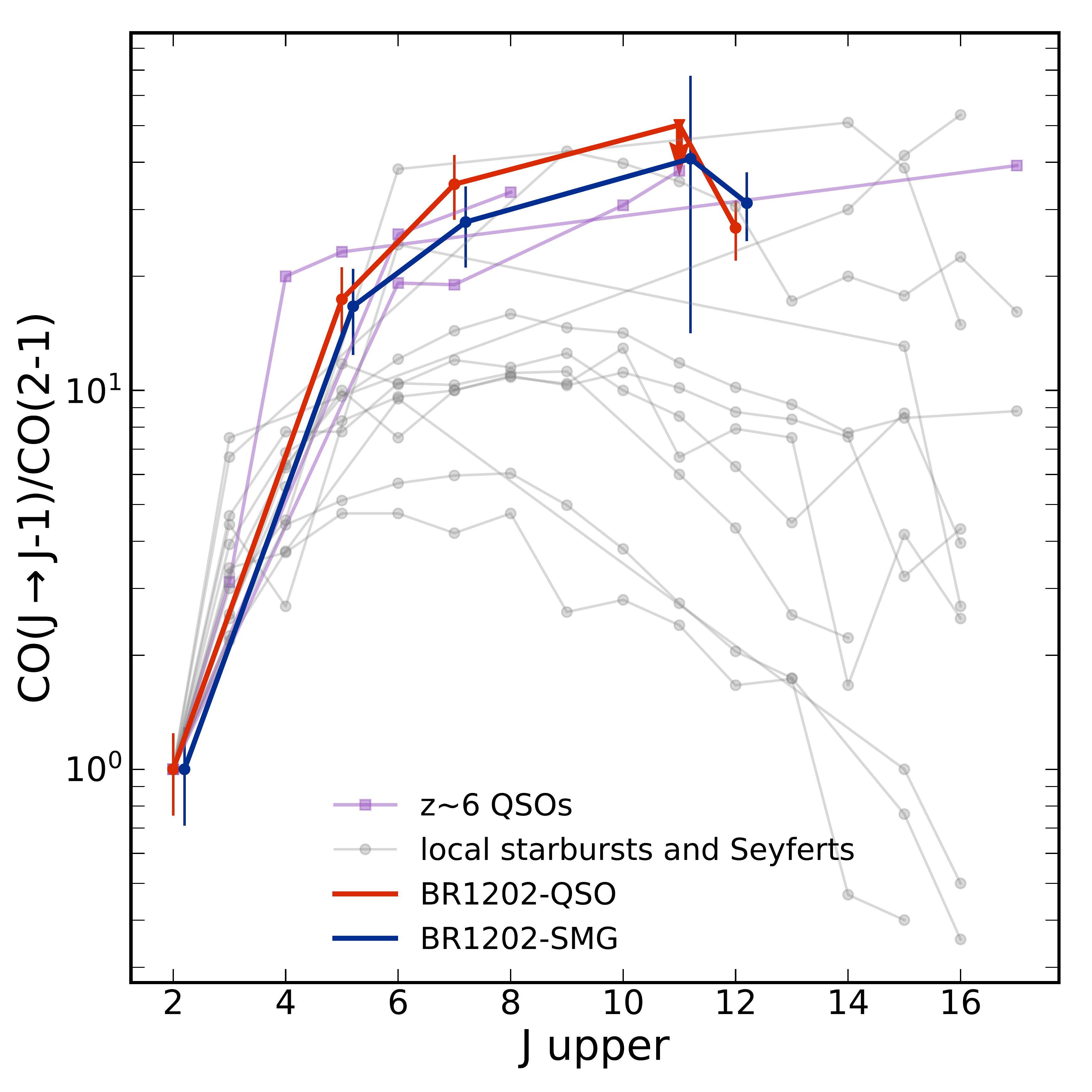}%, bb = 0 0 1024 768 ,
\caption{CO SLEDs for the QSO and the SMG (from this work and \citealt{Salome2012}) relative to $z\sim6$ QSOs (\citealt{Carniani2019}) and local AGNs and starburst galaxies (\citealt{Mashian2015}), where CO~(2--1) luminosities are available for normalization. The CO SLEDs for the SMG is shifted rightward by 0.2 for clarity. \label{fig:cosled}}
\end{figure}

As another, supportive evidence, we report the first detection of CO~(12--11) line in the SMG and the QSO (see Appendix~\ref{app:co1211} for images and the spectra). 
In Figure~\ref{fig:cosled},  we show the CO spectral energy distribution (CO SLED) for various targets from local and high-$z$ galaxy studies (\citealt{Mashian2015,Carniani2019}).
For the CO SLEDs of the BR1202-0725 pair, we took values from \citet{Salome2012} for $J_{\rm upper} = 2,5, 7$, and $11$ ($J_{\rm upper}=2$ values were originally from \citet{Carilli2002b} but corrected for the VLA bandwidth).
While we defer a modeling of the CO SLED, the CO SLEDs of the SMG and the QSO are similar to those of $z\sim6$ QSOs, local Seyferts (NGC~1068, NGC~4945) or warm ULIRGs (NGC~4418).
These similarities provide additional supporting evidence that the QSO and the SMG have similar gas properties of warm and dense gas.

\subsection{Comparison with a cosmological simulation}
Finally, we compare with a cosmological simulation (\citealt{Lupi2020}) which allows us to infer the physical properties from the line ratio. 
They performed a cosmological zoom-in simulation targeting $M_{\rm vir}\sim3\times10^{11}$ $M_{\odot}$ halo at $z=6$ and investigated the far-infrared fine-structure lines.
While the target redshift and galaxies ($M_{\star} \sim 10^{10}$  $M_{\odot}$ and SFR $\approx45$ $M_{\odot}$ yr$^{-1}$ at $z=6$) are different from the star-bursting pair (SFR $\approx 1000 M_{\odot}$ yr$^{-1}$) at $z=4.7$, it provides an insight regarding the \oi/\cii line ratio.
They ran three different models, two (CloudyVAR and CloudyFIX) using \textsc{CLOUDY} (\citealt{Ferland1998}) and one (Krome) \textsc{KROME} (\citealt{Grassi2014}) to test how photoinoization equilibrium and thermal state assumptions affect the FIR emission properties. 
The thermodynamics and chemistry are fully coupled in the Krome model, whereas in the two Cloudy models they are not.
CloudyFIX assume a constant temperature to take into account any dynamical effects while for CloudyVAR the temperature is variable according to the radiation attenuation.
Krome does not take into account the chemical network e.g., CO,  while CloudyVAR and CloudyFIX do.

In the left panel of Figure~\ref{fig:otoc}, we plot the data points from three different models presented in \citet{Lupi2020}.
We converted the simulated SFR into $L_{\rm FIR}$ using the \citet{Kennicutt1998a} recipe to overplot in Figure~\ref{fig:otoc}.
This conversion may contain large uncertainties, depending on the assumptions e.g., IMF, star-formation history, contribution of the old stars.
As noted in \citet{Kennicutt1998a}, the conversion can have uncertainties of $\pm$30\% (starburst galaxies) and the inferred $L_{\rm FIR}$ can result in up to a factor of two to three lower value for normal spiral galaxies.
While the converted $L_{\rm FIR}$ is different by two orders of magnitude compared to the QSO and the SMG, the observed \oi-to-\cii line ratios are consistent with those predicted by the CloudyFIX and Krome models.

\citet{Lupi2020} discussed the origin of the differences between different models. As the \oi line is strongly dependent on the gas temperature, they claimed that the lower \oi-to-\cii line ratio in CloudyVAR can be explained by the lower temperature predicted in the model compared to the others.
Both Krome and CloudyFIX simulations show both the higher temperature and gas density distribution in the (luminosity-weighted) density-temperature phase diagram than CloudyVAR.
Meanwhile, the difference between Krome and CluodyFIX could be coming from the nature (or a caveat) of the Krome model where the chemical network (e.g., CO formation, highly ionized species) is not fully taken into account. 
In the Cloudy models, the calculation is post-processed that all input values are already attenuated than the intrinsic flux that would affect the chemistry. 
Further, in both Cloudy models, gas shocks are not taken into account and a temperature may not be fully consistent with the hydrodynamics in the simulation.
We should note again that the simulated galaxy is different from our galaxies in that they have different galaxy properties (e.g., $M_{\rm star}$, SFR) and redshift. 
As the simulation does not give any information on how dusty galaxy they are, another uncertainty comes in for the conversion between SFR and $L_{\rm FIR}$.  
Despite the different caveats in the models and different galaxy properties between the observed and simulated galaxies, the comparison between the observations and simulations suggest the existence of dense and warm neutral gas in the BR1202-0725 system.\\

Taking all together, the high \oi/\cii ratios in the SMG and the QSO reasonably indicate the existence of dense and warm neutral gas.
The SMG shares ISM properties with the QSO, where the black hole accretion is actively happening.
While, at this moment, it is difficult to conclude whether our observational fact challenges the starburst-QSO evolutionary scenario (e.g., \citealt{Hopkins2008a}), it is tempting to say that the highly obscured starburst (the SMG) and the QSO have similar ISM properties and the SMG might have a hidden AGN, even though both galaxies have not encountered a final coalescence.

\begin{figure}[t]
\centering
\includegraphics[width=0.48\textwidth, bb=0 0 1000 1000]{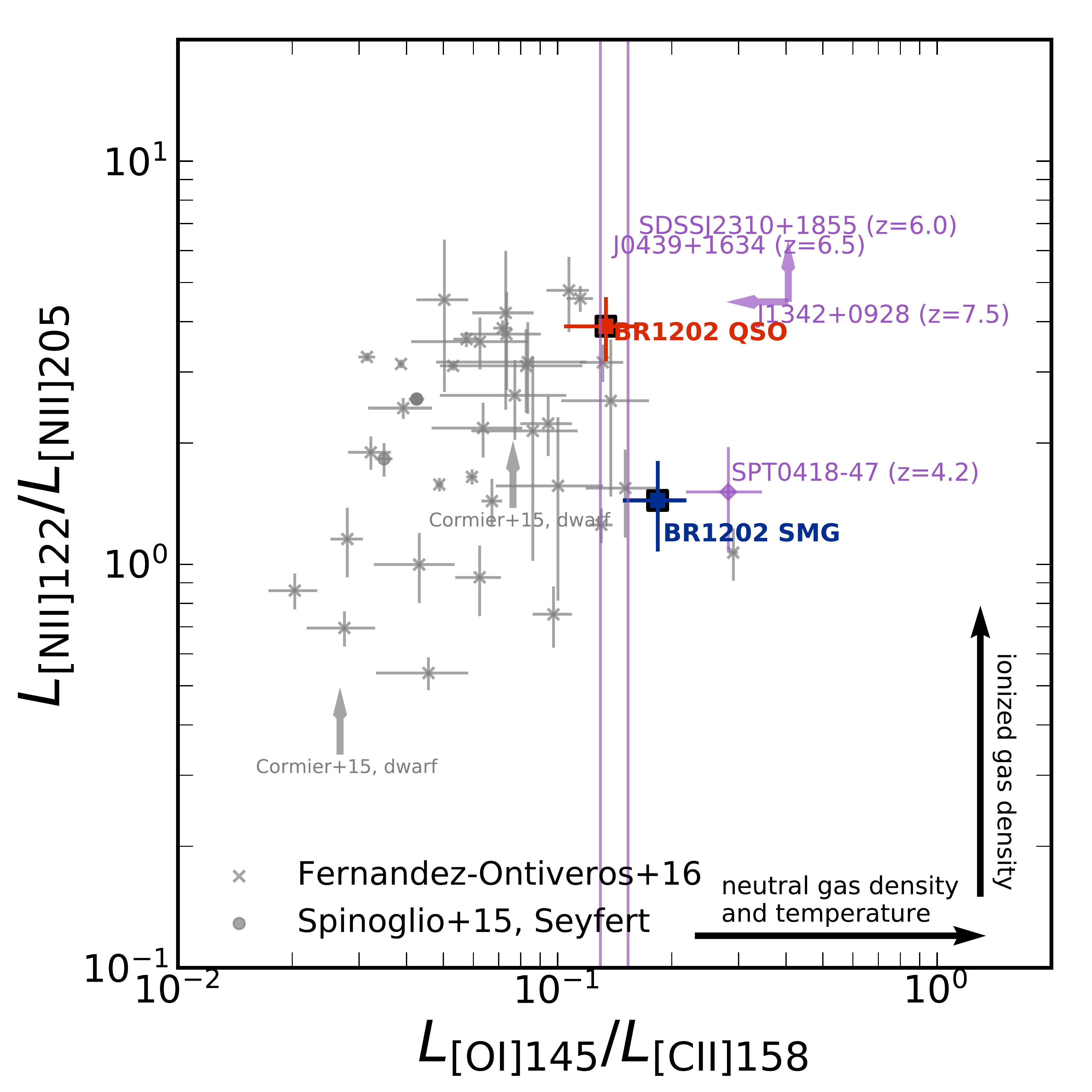}%, bb = 0 0 1024 768 ,
\caption{Luminosity line ratios of \nii$_{122}$/\nii$_{205}$ versus \oi$_{145}$/\cii$_{158}$. Together with our work, we plot local studies from \citealt{FernandezOntiveros2016} (crosses), \citealt{Spinoglio2015} (filled circles), and one SPT galaxy at $z=4.2$ from \citet{DeBreuck2019} having all the four lines detected. Two lower limits of the \nii122/205 ratio from dwarf galaxies from \citet{Cormier2015} are also plotted. We place two lines indicating the \oi/\cii ratio from $z\sim6-7$ QSO observations without \nii$_{122/205}$ constraints (\citealt{Yang2019, Li2020}) and upper limit of the \oi/\cii ratio and lower limit of the \nii ratio from \citet{Novak2019}. \label{fig:ratios}}
\end{figure}

\subsection{Multi-phase properties}
Figure~\ref{fig:ratios} shows the \nii$_{122}$/\nii$_{205}$ versus \oi$_{145}$/\cii$_{158}$ ratios of the QSO and the SMG.
We gathered all data in the literature from local and high-$z$ galaxy studies when they are available (\citealt{Cormier2015, Spinoglio2015, FernandezOntiveros2016, DeBreuck2019, Novak2019, Yang2019, Li2020}).
The figure compares the ionized gas density on the vertical axis versus the neutral gas density and temperature on the horizontal axis.

Two dust-obscured star-forming galaxies (BR1202-SMG and SPT0418-47) have relatively low \nii$_{122}$/\nii$_{205}$ ratios compared to its high \oi/\cii ratios, which are similar to the other high-$z$ samples.
As shortly discussed in \citet{minju2019b}, the optical depth can affect the line fluxes at high frequencies and, in particular, it may lead to a fainter \nii122 flux, thus lowering the \nii ratio.
The \oi and \cii lines are closer in frequency and both of them at lower frequency thus may be less affected than the \nii line ratio.
We also discussed in the previous paper that \nii may not trace very dense ionized gas because a combination of the \nii lines can trace gas density up to $\approx$ 500 cm$^{-3}$. 
In this regard, it may be worth noting on one exceptional data point of NGC~4151, which exhibits a low \nii ratio at the given \oi-to-\cii ratio in the local samples (i.e., an x-cross just below SPT0418-47). 
It is a Seyfert galaxy where its electron density is found to be high ($\approx1700$ cm$^{-3}$) if it is derived based on the [Ne V]. Considering this, the low values in the \nii ratios in these two dust-obscured star-forming galaxies may also indicate the existence of even denser gas that cannot be traced by \nii.

For galaxies detected in all four lines (a total number of 43 including the lower limit constraints from dwarf galaxies), there is a possible positive correlation, but statistically not significant, between the ionized gas density (traced by the \nii ratio) and neutral gas density and temperature (traced by \oi/\cii; Spearman's correlation coefficient = 0.24 with $p$-value 0.13). 
However, if we exclude three `outliers' of BR1202 SMG, SPT0418-47 and NGC~4151 whose \nii ratios are low compared to their high \oi-to-\cii ratio, we get a more significant correlation signature (i.e., Spearman's coefficient of 0.37 with a $p$-value of 0.02).
Of the zeroth order, the positive correlation is naively expected if the \nii$_{122/205}$ and \oi$_{145}$/\cii$_{158}$ ratios trace density of the \hii region and PDR which are physically connected with each other.
By taking into account the correlation between $L_{\rm FIR}$ and \oi/\cii and the one between \nii and \oi/\cii, extreme SFRs for high-$z$ galaxies can be attributed to the existence of dense gas in both phases, ionized and neutral.
We defer more sophisticated models to explain the observed line ratios and ISM structures and conditions to a future article.\\

To conclude, we reported the detection of \oi$_{145}$ from a compact group of BR1202-0725 system at $z=4.7$.
This adds two more galaxies in addition to three in the currently available detection for galaxies at $z>4$.
We find high \oi/\cii ratios compared to local galaxies for all high-$z$ galaxies which exhibit high FIR luminosities.
The high \oi/\cii ratios and the joint analysis with previous detection of \nii lines for both of the QSO and the SMG suggest the presence of warm and dense neutral gas in these highly star-forming galaxies.
The detection of the \oi line in both systems with a relatively short amount of integration with ALMA demonstrates the great potential of this line as a dense gas tracer for high-z galaxies.
Yet, we are still probing highly star-forming exemplars, the \oi line detections of `normal' galaxies are also foreseen in future observations.

\acknowledgments
We thank to the anonymous referee for constructive comments that contributed to the improvement of this work. 
This paper makes use of the following ALMA data: ADS/JAO.ALMA \#2013.1.00745.S, \#2013.1.00259.S. ALMA is a partnership of ESO (representing its member states), NSF (USA) and NINS (Japan), together with NRC (Canada), MOST and ASIAA (Taiwan), and KASI (Republic of Korea), in cooperation with the Republic of Chile. The Joint ALMA Observatory is operated by ESO, AUI/NRAO and NAOJ.
SC acknowledges support from the European Research Council No. 740120 NTERSTELLAR. RM acknowledges ERC Advanced Grant 695671 "QUENCH" and support by the Science and Technology Facilities Council (STFC)
\vspace{20mm}
\facilities{ALMA}
\\

\software{astropy \citep{astropy},  CASA \citep{McMullin2007}
          }



\appendix
\section{Detection of CO(12-11)}\label{app:co1211}
\begin{figure}[b]
\centering
\includegraphics[width=0.48\textwidth, bb=0 0 1000 1000]{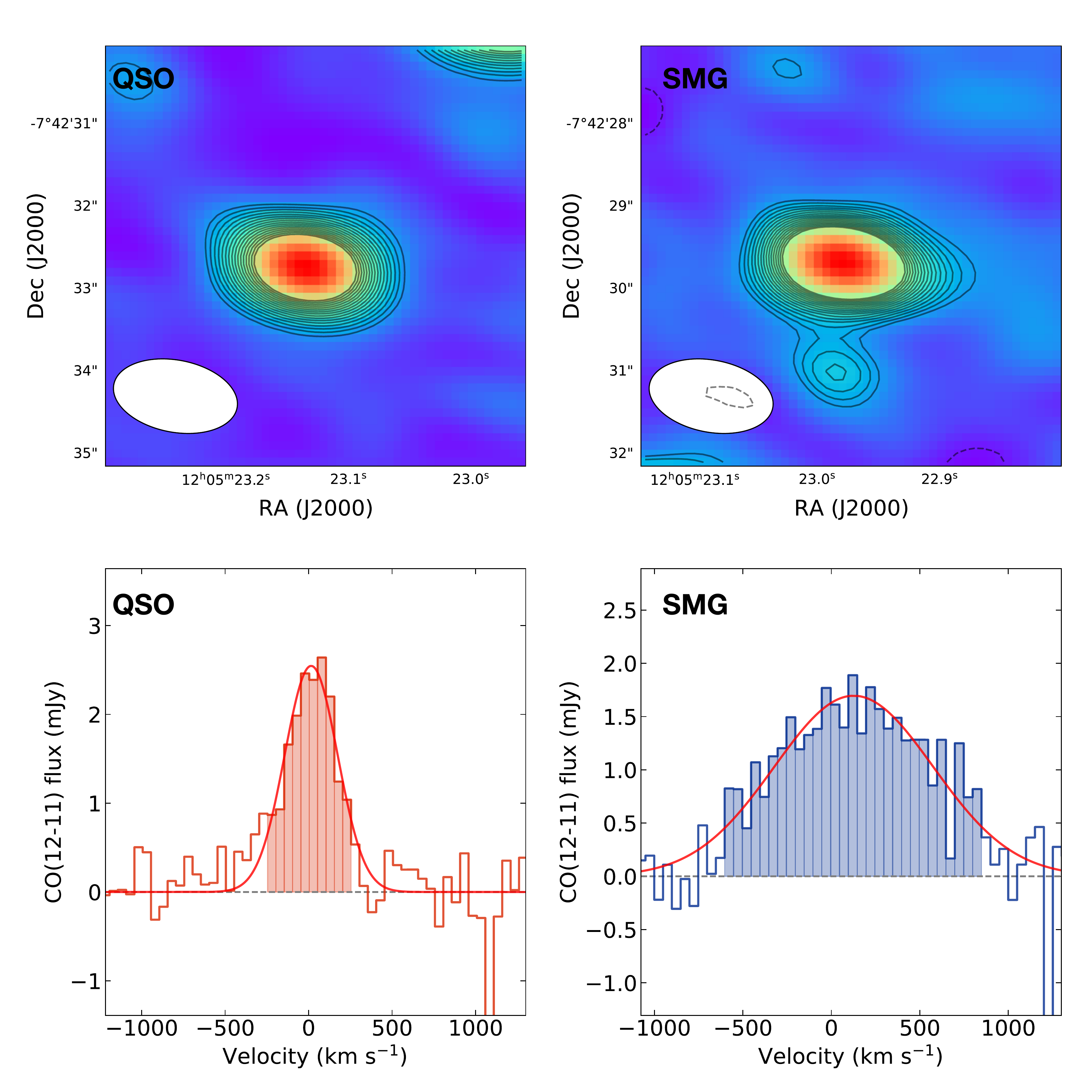}%, bb = 0 0 1024 768 ,
\caption{Upper: CO~(12--11) line maps for QSO (left) and SMG (right). The contours are drawn starting from 4$\sigma$ in steps of 1$\sigma$ up to 10$\sigma$, i.e., [4,5,6, ..., 10]$\sigma$. Negative component contours at -4$\sigma$ is also drawn as dashed lines. The synthesized beam ($1''.52\times0''.88$) is shown on the bottom left on each panel. The panel size is $5''$ in width. Bottom : the CO~(12--11) line spectrum using the same image in the upper panel for the QSO (left) and SMG (right) using a circular aperture of $2''.0$. Shaded area velocity range to create the line intensity map. The spectral resolution is set to 50 km s$^{-1}$. \label{fig:co1211}}
\end{figure}

\begin{figure}[b]
\centering
\includegraphics[width=0.48\textwidth, bb=0 0 1000 1000]{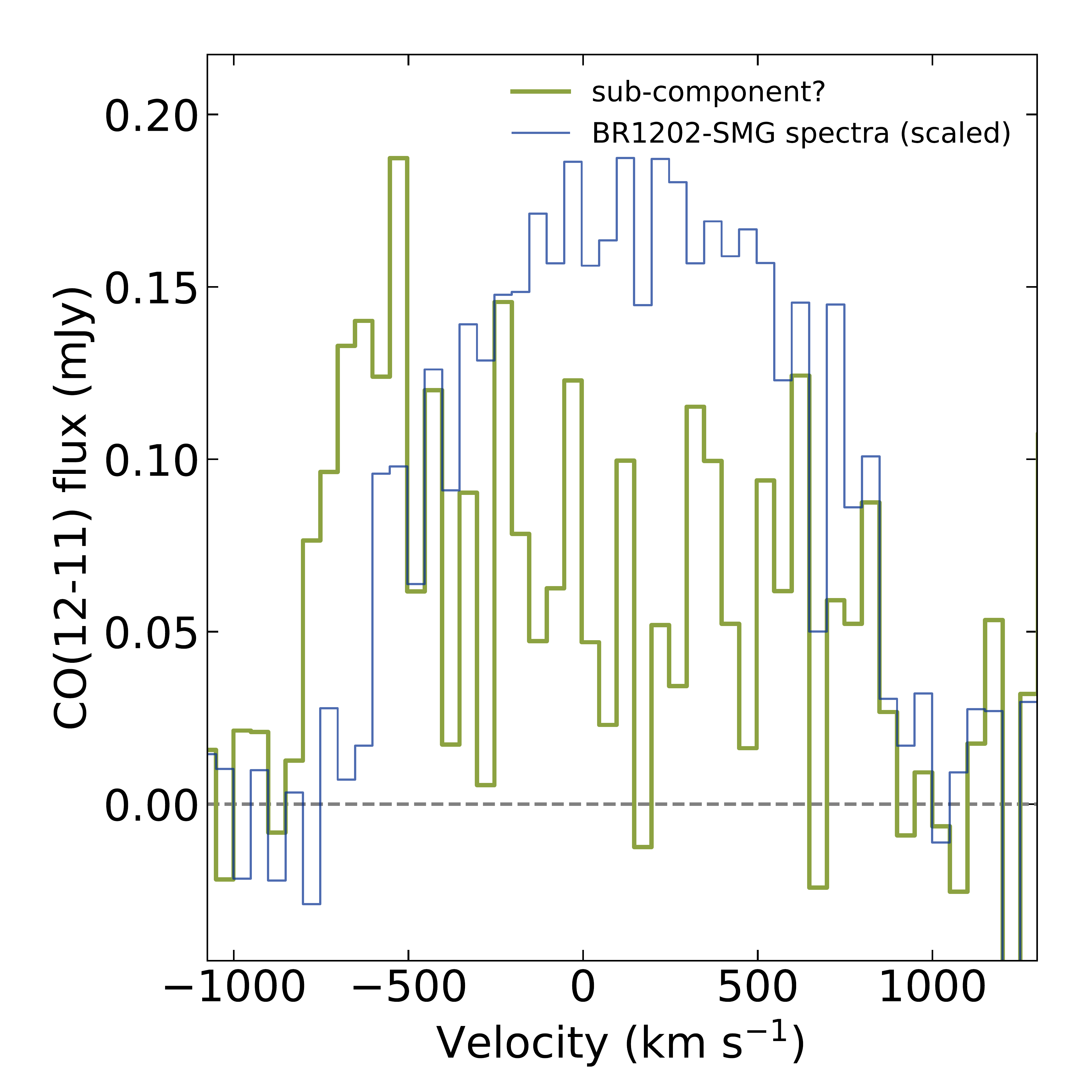}%, bb = 0 0 1024 768 ,
\caption{The spectra at the position of the southern component close to the SMG taking a circular aperture of $0''.5$. The normalized spectra of the SMG (using the same aperture size) is also shown as a reference.\label{fig:subcomp}}
\end{figure}

We detect CO~(12--11) line emissions both from the SMG and the QSO. Figure~\ref{fig:co1211} shows the integrated line maps and the spectrum. 
Three LAEs are not detected in CO~(12--11).
The line fluxes, FWHM and luminosities are summarized in Table~\ref{tab:lineprop}.
For the SMG, it seems that there is a sub-component associated to it in the southern area which is peaked at $\sim -500$ km s$^{-1}$ with a long tail redward (Figure~\ref{fig:subcomp}).
We could not identify a similar signature in the \cii map, though if it is real, it may be relevant to the \cii~bridge component connected with the QSO (\citealt{Carilli2013b}). 
LAE~1 is known to have an offset between the Ly$\alpha$ and the [CII] emissions, which is $0''.6$. 
Accordingly, we have checked whether the CO~(12--11) subcomponent is associated with the Ly$\alpha$ emission instead of the \cii emission. 
The CO~(12--11) subcomponent is offset from the \cii emission by $1''.1$, which is larger than the Ly$\alpha$--\cii separation. 
Therefore, if the emission is real, this might not be directly associated with the Ly$\alpha$ emission from LAE~1, but with a halo or outflowing component from the SMG, if any. 
The extended Ly$\alpha$ emissions (\citealt{Drake2020}) reach out to a region where the subcomponent is detected. 
Future, deeper observations would need to verify this.

\section{HCN~(6--5) and HCO$^{+}$(6--5) observations}\label{app:hcn}
\begin{figure}[tb]
\centering
\includegraphics[width=0.48\textwidth, bb=0 0 1000 500]{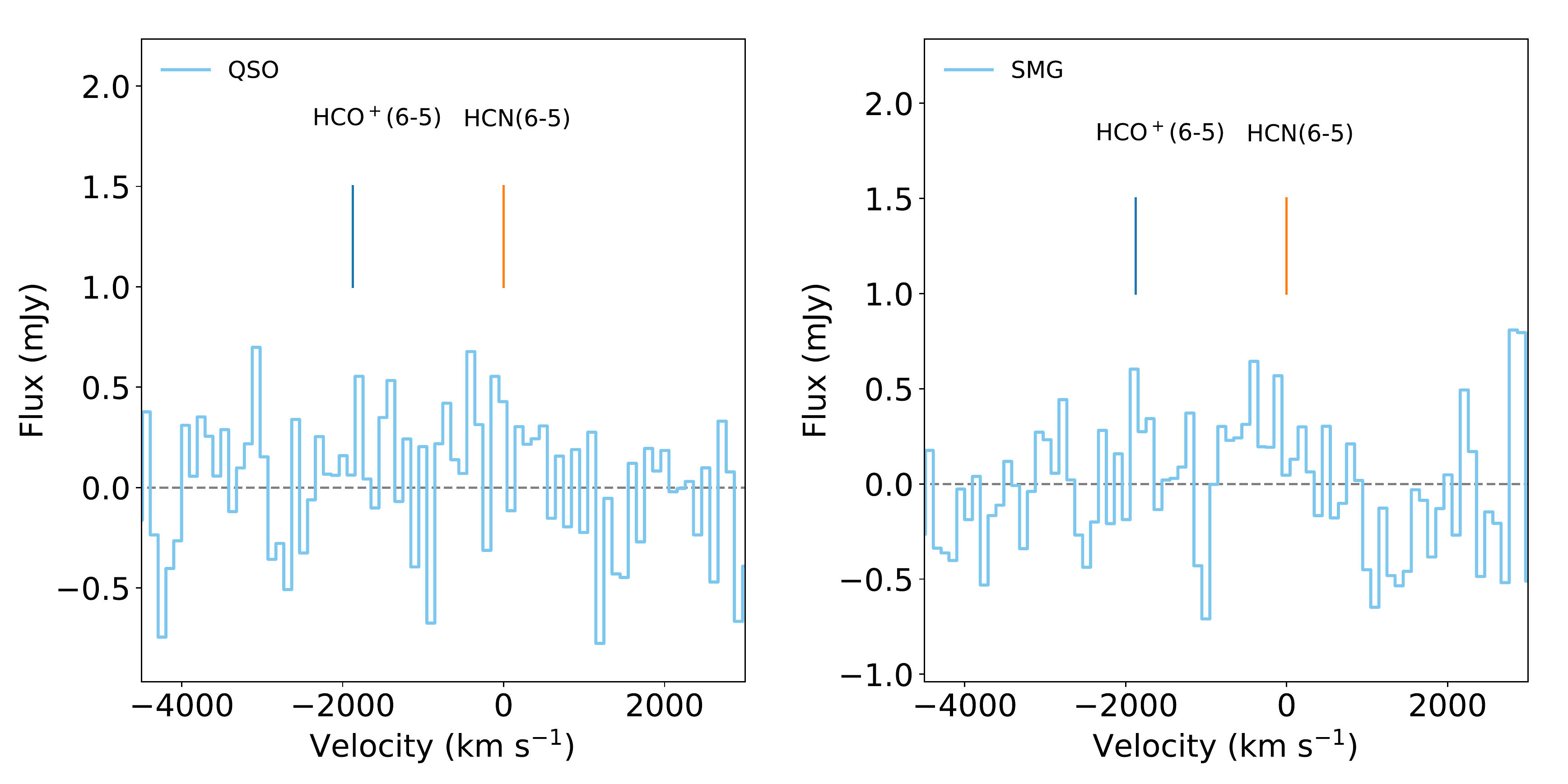}%, bb = 0 0 1024 768 ,
\caption{Non-detection spectrum of HCN~(6--5) and HCO$^{+}$(6--5) lines for the QSO (left) and the SMG (right).\label{fig:hcnhco}}
\end{figure}

At a resolution of $0''.8\times0''.6$, there is no clear detection feature around the expected velocity range (Figure~\ref{fig:hcnhco}).
The spectrum is after the continuum subtraction using $\mathtt{imcontsub}$ (fitorder = 1\footnote{Changing to fitorder = 0 does not change the result.}).
For a circular aperture of $1''.0$, the 3$\sigma$ upper limit placed by this ALMA observations are $<0.27$ Jy km s$^{-1}$and $< 0.76$ Jy km s$^{-1}$, for the QSO and the SMG, respectively, assuming the same line widths measured from the CO~(12-11) emission (Table~\ref{tab:lineprop}) and using the noise level measured at 100 km s$^{-1}$ resolution.

%A handy "cheat sheet" that provides the necessary LaTeX to produce 17 
%different types of tables is available at \url{http://journals.aas.org/authors/aastex/aasguide.html#table_cheat_sheet}.


\bibliography{minjujournal_v2}
\bibliographystyle{aasjournal}



\end{document}